\documentclass[
twocolumn,
amsmath,amssymb,
nofootinbib,
showkeys,
prd,
aps,
10pt,
longbibliography,preprintnumbers
]{revtex4-2}
\pdfoutput=1
\usepackage{hyperref}
\usepackage{xcolor,graphicx}
\usepackage{bm}

\usepackage{soul}\usepackage[normalem]{ulem}

\begin{document}
\interfootnotelinepenalty=10000

\title{Equivalence principles in WTG}

\author{Ana Alonso-Serrano}
\email{ana.alonso.serrano@aei.mpg.de}
\affiliation{Institut für Physik, Humboldt-Universität zu Berlin, Zum Großen Windkanal 6, 12489 Berlin, Germany}
\affiliation{Max-Planck-Institut f\"ur Gravitationsphysik (Albert-Einstein-Institut), \\Am M\"{u}hlenberg 1, 14476 Potsdam, Germany}
	
\author{Luis J. Garay}
\email{luisj.garay@ucm.es}
\affiliation{Departamento de F\'{i}sica Te\'{o}rica and IPARCOS, Universidad Complutense de Madrid, 28040 Madrid, Spain}
	
\author{Marek Li\v{s}ka,}
\email{liskama4@stp.dias.ie}
\affiliation{School of Theoretical Physics, Dublin Institute for Advanced Studies, 10 Burlington Road, Dublin 4, Ireland}
\affiliation{Institute of Theoretical Physics, Faculty of Mathematics and Physics, Charles University, V Hole\v{s}ovi\v{c}k\'{a}ch 2, 180 00 Prague 8, Czech Republic}

\begin{abstract}
There exist two consistent theories of massless, self-interacting gravitons, which differ by their local symmetries: GR and WTG. We show that these two theories are also the only two metric descriptions of gravity in $4$ spacetime dimensions which obey the equivalence principle for test gravitational physics. We further analyse how the weaker formulations of the equivalence principle are realised in WTG (and its generalisations). The analysis sheds light on the behaviour of matter fields in this theory.
\end{abstract}

\preprint{IPARCOS-UCM-25-011}

\maketitle
\tableofcontents

\section{Introduction}

The equivalence principle (EP) historically played an important role in guiding the development of GR~\cite{Einstein:1911}. It essentially states that suitably defined ``test physics'' is not affected by the presence of a background gravitational field~\cite{Casola:2015,Fletcher:2022a,Fletcher:2022b}. In this context, test physics refers to localised physical processes whose effect on the background gravitational field is negligible. There exists a hierarchy of different formulations of the EP, depending on which processes one chooses as admissible test physics. For instance, the weak EP only concerns geodesic motion of test particles with negligible self-gravity. The most restrictive version known as the strong EP extends to arbitrary test physics including gravitational phenomena, such as the propagation of gravitational waves (provided that their backreaction on the background spacetime can be neglected). The classification of the various formulations of the EP we adopt follows Ref.~\cite{Casola:2015}. In general, we can separate the EPs into two groups. The ones concerning non-gravitational test physics constrain the interaction between matter and fixed gravitational background, essentially determining the kinematics of gravity~\cite{Fletcher:2022a,Fletcher:2022b}. The EPs which extend to gravitational test physics then also impose constraints on gravitational dynamics compatible with them~\cite{Casola:2014}.
 
Nowadays, the interest in the EP lies in the experimental and observational studies of its validity~\cite{Overduin:2009,Liberati:2013,Will:2014} and in studies of its validity in quantum physics~\cite{Giacomini:2020,Wagner:2023,Balsells:2023,Zhang:2024}. Moreover, the various candidate theories of gravity can be classified according to which formulations of the EP they are compatible with~\cite{Casola:2014,Casola:2015}. In particular, the only theory known to obey the strong EP is general relativity (GR).

Herein, we show that the strong EP is compatible with at least one other gravitational theory in $4$ dimensions, namely, with Weyl transverse gravity (WTG)~\cite{Unruh:1989,Finkelstein:2001,Alvarez:2006,Alvarez:2013,Alvarez:2013b,Barcelo:2014,Carballo:2015,Alvarez:2016,Barcelo:2018,Carballo:2020,Carballo:2022,Garay:2023,Garcia:2023}. This theory has the same classical solutions as GR\footnote{Note here that explicitly WTG solutions are equivalent to the whole set of GR theories, one for each value of the cosmological constant, because in this theory the cosmological constant emerged as a constant of motion which can take different values while in GR the cosmological constant is fixed upon choosing its value in the action.}. However, its equations of motion are traceless and the theory has different local symmetries. Rather than being invariant under arbitrary diffeomorphisms (Diff-invariant), it is instead invariant under the subgroup of spacetime volume preserving diffeomorphisms and under Weyl transformations (WTDiff-invariant). Both theories thus have $D\left(D+1\right)/2$ local symmetries, with $D$ being the spacetime dimension. WTG first emerged from the construction of a theory of massless, self-interacting gravitons~\cite{Alvarez:2006,Barcelo:2014,Alvarez:2016,Barcelo:2018,Carballo:2022}. There exist precisely two such theories that do not involve any gauge fixing and their only propagating degrees of freedom are those of the graviton. These theories differ by the choice of the symmetry group, with Diff invariance leading to GR and WTDiff invariance to WTG.

WTG passes all the experimental and observational tests up to date (having the same classical solutions as GR), provides a consistent theory for self-interacting gravitons, and offers a robust solution for one of the problems related to the value of the  cosmological constant \cite{Unruh:1989,Carballo:2015,Barcelo:2018}. Thence, it represents a viable and relevant alternative to GR. By showing that WTG satisfies the strong EP, we offer a further argument for singling it out as a competitor of GR.

In addition, WTG is actually favoured over GR by approaches that derive the gravitational dynamics from thermodynamic considerations~\cite{Alonso:2024} (for a broader context on such thermodynamic derivations see, e.g.~\cite{Jacobson:1995ab,Padmanabhan:2010,Jacobson:2015}). Since one of the key assumptions of any local thermodynamic derivation is the validity of the strong EP~\cite{Chirco:2010}, checking that this assumption is consistent with WTG provides an additional motivation for this work.

Aside from studying the strong EP, we also discuss various weaker formulations of the EP and show how they are incorporated in WTG. Furthermore, we study their validity even in more general theories of gravity with WTDiff symmetry.

The paper is organised as follows. In section~\ref{WTG}, we review the basics of WTG and of more general WTDiff-invariant theories. Section~\ref{NG EP} discusses the EPs restricted to non-gravitational test physics. We then consider the EPs for gravitational test physics in section~\ref{G EP}. Finally, section~\ref{conclusion} summarises our findings and discusses possible future development.

We work in an arbitrary spacetime dimension $D$ (unless specified otherwise) with a metric signature $(-,+,...,+)$. We set $c=k_{\text{B}}=G=\hbar=1$. Lowercase Greek letters are used for abstract spacetime indices. Other conventions follow~\cite{MTW}.

\section{WTG and its generalisations}
\label{WTG}

Herein, we recall the main features of WTG and of WTDiff-invariant theories of gravity in general. For a detailed review, we refer the readers to~\cite{Carballo:2022}.

First of all, to construct any WTDiff-invariant theory of gravity, one needs to introduce a non-dynamical volume $D$-form, $\boldsymbol{\omega}=\omega\left(x\right)\text{d}x^{0}\wedge\text{d}x^{1}\wedge...\wedge\text{d}x^{D-1}$, where $\omega$ is a strictly positive scalar density~\cite{Carballo:2015,Barcelo:2018}. The definition of the spacetime then includes, aside from the usual manifold $\mathcal{V}$ and the metric $g_{\mu\nu}$, also the conformal structure given by the volume $D$-form $\boldsymbol{\omega}$, i.e., $\left(\mathcal{V},g_{\mu\nu},\boldsymbol{\omega}\right)$.

To simplify the notation, we define an auxiliary, WTDiff-invariant metric constructed from the dynamical metric $g_{\mu\nu}$ and the background volume measure $\omega$,
\begin{equation}
\label{aux metric}
\tilde{g}_{\mu\nu}=\left(\sqrt{-\mathfrak{g}}/\omega\right)^{-2/D}g_{\mu\nu},
\end{equation}
where $\mathfrak{g}$ denotes the metric determinant. Both $\sqrt{-\mathfrak{g}}$ and $\omega$ are scalar densities of weight $+1$, ensuring that $\tilde{g}_{\mu\nu}$ (which depends on their ratio) is a tensor. This auxiliary metric can be understood as a restriction of $g_{\mu\nu}$ to the unimodular gauge, $\sqrt{-\mathfrak{g}}=\omega$. To ensure we always work with WTDiff-invariant expressions, raising and lowering of indices is performed with $\tilde{g}_{\mu\nu}$ and its inverse metric $\tilde{g}^{\mu\nu}$. We stress that we keep $g_{\mu\nu}$ as the dynamical field. However, $\tilde{g}_{\mu\nu}$ plays the role of our kinematical metric to which the matter fields couple. Thence, as we discuss in more detail in the next section, we also consider all the measurements of distances, time intervals, angles etc., with respect to $\tilde{g}_{\mu\nu}$. At the level of classical physics, one might then forget about $g_{\mu\nu}$ altogether and work solely with $\tilde{g}_{\mu\nu}$. However, the fixed gauge $\sqrt{-\mathfrak{g}}=\omega$ for the determinant of $\tilde{g}_{\mu\nu}$ has been shown to lead to problems on the level of the effective quantum theory~\cite{Padilla:2014,Alvarez:2023a}. No such problems appear when one defines the graviton with respect to the unconstrained metric $g_{\mu\nu}$~\cite{Carballo:2015,Carballo:2022,Martin:2024}. Hence, keeping $g_{\mu\nu}$ as the dynamical field appears to be necessary for a consistent quantisation of the theory. That said, our results in this paper are fully classical and, therefore, they also straightforwardly apply to the case when one considers the gauge fixed metric $\tilde{g}_{\mu\nu}$ as the dynamical field.

The Levi-Civita connection defined with respect to $\tilde{g}_{\mu\nu}$ reads
\begin{equation}
\label{connection}
\tilde{\Gamma}^{\mu}_{\;\:\nu\rho}=\Gamma^{\mu}_{\;\:\nu\rho}-\frac{1}{D}\left(\delta^{\mu}_{\nu}\delta^{\lambda}_{\rho}+\delta^{\mu}_{\rho}\delta^{\lambda}_{\nu}-g_{\nu\rho}g^{\lambda\mu}\right)\partial_{\lambda}\ln\frac{\sqrt{-\mathfrak{g}}}{\omega},
\end{equation}
where $\Gamma^{\mu}_{\;\:\nu\rho}$ denotes the connection which is Levi-Civita with respect to the dynamical metric, $g_{\mu\nu}$. 

For the dynamical metric $g_{\mu\nu}$ it holds
\begin{equation}
\tilde{\nabla}_{\rho}g_{\mu\nu}=g_{\mu\nu}\frac{2}{D}\nabla_{\rho}\ln\frac{\sqrt{-\mathfrak{g}}}{\omega}.
\end{equation}
Since the result is a tensor product of the metric and a gradient of a scalar function, it follows that $\tilde{\Gamma}^{\mu}_{\;\:\nu\rho}$ is an integrable Weyl connection for the dynamical metric, as well as for any metric conformally related to it (for details on Weyl connections see, e.g.~\cite{Folland:1970}). This observation will play a role in our discussion of the EPs.

Using the Weyl connection~\eqref{connection}, we introduce an auxiliary, WTDiff-invariant Riemann tensor
\begin{equation}
\label{Riemann}
\tilde{R}^{\mu}_{\;\:\nu\rho\sigma}= 2\tilde{\Gamma}^{\mu}_{\;\:\nu[\sigma,\rho]}+2\tilde{\Gamma}^{\mu}_{\;\:\lambda[\rho}\tilde{\Gamma}^{\lambda}_{\;\:\sigma]\nu}.
\end{equation}

The simplest action one can construct from the auxiliary metric and the corresponding Riemann tensor is that of WTG, i.e.,
\begin{equation}
\label{I WTG}
I_{\text{WTG}}=\frac{1}{16\pi G}\int_{\text{V}}\left(\tilde{R}+L_{\psi}\right)\omega\text{d}^{D}x,
\end{equation}
where $\text{V}$ is the spacetime manifold and $\tilde{R}=\tilde{g}^{\mu\nu}\tilde{R}_{\mu\nu}$ denotes the scalar curvature defined with respect to $\tilde{g}_{\mu\nu}$. The Lagrangian $L_{\psi}$ for minimally coupled matter fields is constructed from the matter variables, the auxiliary metric,and the partial derivatives.

Since the spacetime volume measure is non-dynamical, adding any constant term to the WTG Lagrangian amounts simply to shifting the action by a constant and does not affect dynamics in any way. Hence, we are free to set this constant term to zero in the following. This marks a departure from GR, where a constant term in the Lagrangian corresponds to the cosmological constant.

By construction, $I_{\text{WTG}}$ is invariant under Weyl transformations
\begin{equation}
g_{\mu\nu}\quad\to\quad e^{2\sigma}g_{\mu\nu},
\end{equation}
or, infinitesimally,
\begin{equation}
\delta g_{\mu\nu}=2\sigma g_{\mu\nu},
\end{equation}
where $\sigma$ is an arbitrary scalar function. We stress that we define the Weyl transformations in the context of Weyl transverse gravity as gauge transformations acting solely on the metric~\cite{Carballo:2022}. The volume measure $\omega$ and the matter fields are then by definition unaffected by Weyl transformations\footnote{As an aside, one may of course consider scale transformations in Weyl transverse gravity that act both on the metric and on the matter fields, much like in general relativity. While these are usually also called ``Weyl transformations'' in the literature on diffeomorphism-invariant gravity, they are distinct from Weyl transformations we discuss here. In particular, these scale transformations are not local symmetries of Weyl transverse gravity (just as they are not local symmetries of general relativity).}, ensuring the Weyl invariance of $\tilde{g}_{\mu\nu}$. Furthermore, $I_{\text{WTG}}$ is invariant under transverse diffeomorphisms, but not under longitudinal ones. However, we must be careful to consider the appropriate notion of transversality. The usual condition on the generator $\xi^{\mu}$ of transverse diffeomorphisms, $\nabla_{\mu}\xi^{\mu}=0$, is not Weyl invariant. Thus, it cannot be satisfied in every Weyl frame simultaneously, making it unsuitable for WTG. Instead, one must define transversality with respect to the Weyl invariant covariant derivative. Hence, the appropriate transversality condition reads
\begin{equation}
\tilde{\nabla}_{\mu}\xi^{\mu}=0\qquad\iff\qquad\nabla_{\mu}\xi^{\mu}=\xi^{\mu}\partial_{\mu}\ln\frac{\sqrt{-\mathfrak{g}}}{\omega}.
\end{equation}
Since the Lie derivative of the volume $D$-form $\boldsymbol{\omega}$ yields $\pounds_{\xi}\boldsymbol{\omega}=\boldsymbol{\omega}\tilde{\nabla}_{\mu}\xi^{\mu}$ (this result can be obtained by direct computation and also follows from the fact that \mbox{$\tilde{\nabla}_{\mu}\boldsymbol{\omega}=0$}), we can understand the transversality condition as defining the volume preserving transformations. Transverse diffeomorphisms transform the dynamical metric $g_{\mu\nu}$ in the usual way
\begin{equation}
\delta_{\xi}g_{\mu\nu}=2\nabla_{(\nu}\xi_{\mu)},
\end{equation}
and they act on the auxiliary metric as
\begin{equation}
\delta_{\xi}\tilde{g}_{\mu\nu}=2\tilde{\nabla}_{(\nu}\xi_{\mu)}.
\end{equation}

To find equations of motion for WTG, we vary action~\eqref{I WTG} with respect to the dynamical metric $g^{\mu\nu}$. Variation of the auxiliary Ricci scalar term yields, up to total divergence terms, the contribution $\left(\sqrt{-\mathfrak{g}}/\omega\right)^{-2/D}\omega\left(\tilde{R}_{\mu\nu}-\tilde{R}\tilde{g}_{\mu\nu}/D\right)$, where the second term comes from the variation of the factor $\left(\sqrt{-\mathfrak{g}}/\omega\right)^{-2/D}$ with respect to $g_{\mu\nu}$. Variation of the matter action similarly yields $\left(\sqrt{-\mathfrak{g}}/\omega\right)^{-2/D}\omega\left(\tilde{T}_{\mu\nu}-\tilde{T}\tilde{g}_{\mu\nu}/D\right)$, where we define the WTDiff-invariant energy-momentum tensor
\begin{equation}
\tilde{T}_{\mu\nu}=-2\frac{\partial L_{\psi}}{\partial\tilde{g}^{\mu\nu}}+L_{\psi}\tilde{g}_{\mu\nu}.
\end{equation}
Dividing by the common factor $\left(\sqrt{-\mathfrak{g}}/\omega\right)^{-2/D}\omega$, we  obtain the traceless, WTDiff-invariant equations of motion
\begin{equation}
\label{EoMs}
\tilde{R}_{\mu\nu}-\frac{1}{D}\tilde{R}\tilde{g}_{\mu\nu}=8\pi G\left(\tilde{T}_{\mu\nu}-\frac{1}{D}\tilde{T}\tilde{g}_{\mu\nu}\right).
\end{equation}

Diff-invariance of gravitational dynamics implies directly that $\nabla_{\nu}T_{\mu}^{\;\:\nu}=0$ and, consequently, that energy-momentum is locally conserved. However, this is not generally the case for WTDiff-invariant theories, since invariance under transverse diffeomorphisms only leads to a weaker condition~\cite{Alvarez:2013}
\begin{equation}
\label{div T}
\tilde{\nabla}_{\nu}\tilde{T}_{\mu}^{\;\:\nu}=\tilde{\nabla}_{\mu}\mathcal{J},
\end{equation}
where $\mathcal{J}$ is a scalar function which quantifies the local energy-momentum non-conservation~\cite{Alonso:2021,Carballo:2022}. 

Bianchi identities allow us to restate the traceless equations of motion~\eqref{EoMs} in a divergenceless form reminiscent of the equations of motion of GR
\begin{equation}
\label{Einstein-style}
\tilde{R}_{\mu\nu}-\frac{1}{2}\tilde{R}\tilde{g}_{\mu\nu}+\Lambda\tilde{g}_{\mu\nu}=8\pi G \tilde{T}'_{\mu\nu},
\end{equation}
where $\Lambda$ is an arbitrary integration constant. This represents the principal difference from general relativity, where the cosmological constant is an (arbitrary) fixed parameter in the gravitational action. The divergenceless tensor
\begin{equation}
\label{T divergenceless}
\tilde{T}'_{\mu\nu}=\tilde{T}_{\mu\nu}-\mathcal{J}\tilde{g}_{\mu\nu},
\end{equation}
acts as the matter source for the Einstein tensor. Hence, $\tilde{T}'_{\mu\nu}$ is the relevant tensor for coupling of matter fields to WTG, as we will see in our discussion of the EP. Let us note that, given the traceless part of the energy-momentum tensor, the corresponding divergenceless tensor $\tilde{T}'_{\mu\nu}$ is defined uniquely up to a shift by a term of the form $C\tilde{g}_{\mu\nu}$, with $C$ being an arbitrary constant. Since this constant shift can be absorbed into the arbitrary integration constant $\Lambda$, $\tilde{T}'_{\mu\nu}$ is completely fixed by the traceless gravitational equations of motion.

The integration constant $\Lambda$ plays the role of the cosmological constant. In contrast to GR, $\Lambda$ is unrelated to any parameter of the Lagrangian and is only defined on shell, generically taking different values for various solutions of the equations of motion. In other words, $\Lambda$ represents a global degree of freedom of the theory. 

These WTDiff-invariant theories belong to the so called unimodular theories of gravity. Note that fixing the gauge in WTG will bring us to unimodular gravity theory, that is not longer invariant under Weyl transformations \cite{Gielen:2018pvk,Garcia:2023}. These scenarios can be useful for example in cosmology, although in other more fundamental works one faces losing some information and consistency~\cite{Padilla:2014,Alvarez:2023a,Martin:2024}, as it would happen if one considers a similar gauge fixing in electrodynamics.

In the same way that there exist modified Diff-invariant theories of gravity, there exist WTDiff-invariant theories generalising WTG. Their Lagrangian can  be any scalar constructed from the auxiliary metric, $\tilde{g}_{\mu\nu}$, the auxiliary Riemann tensor, $\tilde{R}^{\mu}_{\;\:\nu\rho\sigma}$, the Weyl covariant derivative $\tilde{\nabla}_{\mu}$ and some collection of matter fields, in principle non-minimally coupled. One can in fact show that for each local, WTDiff-invariant action there exists a corresponding local, Diff-invariant action which implies the same classical gravitational dynamics (except for the different behaviour of $\Lambda$) and vice versa~\cite{Carballo:2022}. In other words, there exist pairs of Diff and WTDiff-invariant theories of gravity that are mutually equivalent in the same sense in which GR is equivalent to WTG. As a special case, there exist WTDiff-invariant Lanczos-Lovelock theories~\cite{Lanczos:1938,Lovelock:1971,Padmanabhan:2013} which are the only purely metric, WTDiff-invariant theories with second order equations of motion. They have the same classical solutions as the corresponding Diff-invariant Lanczos-Lovelock gravitational theories~\cite{Padmanabhan:2013}. As we comment in section~\ref{G EP}, Lanczos-Lovelock gravity, both Diff- and WTDiff-invariant, is also notable for obeying the strong equivalence principle.

\subsection{Example: universe filled with a barotropic perfect fluid}

To conclude this section, we demonstrate the key features of Weyl transverse gravity on the instructive example of a homogeneous, isotropic, spatially flat universe filled with a barotropic perfect fluid with an equation of state $p=\left(\gamma-1\right)\rho$ where $\gamma\in\left[1,2\right]$. The metric we consider reads
\begin{equation}
\text{d}s^2=-\text{d}t^2+a^2\left(t\right)\text{d}x_i\text{d}x^i,
\end{equation}
where $a\left(t\right)$ is the scale factor. The traceless part of the WTDiff-invariant energy-momentum tensor of the perfect fluid reads
\begin{equation}
\label{fluid traceless}
\tilde{T}_{\mu\nu}-\frac{1}{D}\tilde{T}\tilde{g}_{\mu\nu}=\left(\rho+p\right)\left(\tilde{u}_{\mu}\tilde{u}_{\nu}+\frac{1}{D}\tilde{g}_{\mu\nu}\right),
\end{equation}
where $\tilde{u}_{\mu}$ the (WTDiff-invariant) velocity of the fluid. The traceless equations of motion~\eqref{EoMs} yield only one independent equation (compared to two equations of general relativity)
\begin{equation}
\label{Raych}
\dot{H}=-4\pi\left(\rho+p\right)=-4\pi\gamma\rho,
\end{equation}
where the overdot denotes a time derivative and $H=\dot{a}/a$ is the Hubble rate. The right hand side of this equation depends only on the combination $\rho+p$, not separately on $\rho$ and $p$ as for general relativity. The divergence condition~\eqref{div T} then provides the second independent equation
\begin{equation}
\gamma\dot{\rho}+3H\rho=-\dot{\mathcal{J}}.
\end{equation}
If $\mathcal{J}=0$, we obtain $\rho=\rho_0/a^{3\gamma}$, where $\rho_0$ is an integration constant. Plugging this result into equation~\eqref{Raych}, multiplying it by $H$, and integrating, then gives us the Friedmann equation
\begin{equation}
H^2=\frac{8\pi}{3}\rho+\frac{\Lambda}{3},
\end{equation}
where $\Lambda$ is an integration constant. Then, we precisely recover the dynamics of general relativity, except for the above discussed different origin of the cosmological constant $\Lambda$. If $\mathcal{J}\ne0$, we can simply switch to the divergenceless energy-momentum tensor $\tilde{T}'_{\mu\nu}$ given by equation~\eqref{T divergenceless}. For our perfect fluid, it reads
\begin{equation}
\tilde{T}'_{\mu\nu}=\left(\rho+p\right)\tilde{u}_{\mu}\tilde{u}_{\nu}+\left(p-\mathcal{J}\right)\tilde{g}_{\mu\nu}.
\end{equation}
This tensor then corresponds to pressure $p-\mathcal{J}$ and energy density $\rho+\mathcal{J}$. Since the combination $\rho+p$ is unchanged, the traceless part of this energy-momentum tensor is the same as of the original $\tilde{T}_{\mu\nu}$ (see equation~\eqref{fluid traceless}). Therefore, on the level of the gravitational dynamics of Weyl transverse gravity, one cannot tell apart $\tilde{T}'_{\mu\nu}$ and $\tilde{T}_{\mu\nu}$ as only the traceless part of the energy-momentum tensor is relevant. In this case, we recover the dynamics equivalent to general relativity, coupled to the divergenceless energy-momentum tensor $\tilde{T}'_{\mu\nu}$. Although the case we discussed is very simple, the same qualitative picture holds in generic spacetimes and for arbitrary matter contents.

\section{Equivalence principles for non-gravitational test physics}
\label{NG EP}

We start our discussion of the EPs in WTDiff-invariant gravity with the formulations that do not consider gravitational test physics. These EPs constrain the way in which matter interacts with the fixed background gravitational field rather than directly the gravitational dynamics (in other words, they concern the kinematics of the gravitational field). Therefore, they can be incorporated into any local Diff-invariant theory of gravity, provided that matter fields obey the necessary conditions~\cite{Fletcher:2022b}. We shall see that the same statement applies to WTDiff-invariant theories, although the analysis of the geodesic motion requires some care. We proceed by checking the validity of the three different EPs for non-gravitational test physics one by one.

\subsection{Newton equivalence principle}

Before going to the more complicated relativistic setting, we first briefly address the weakest formulation of the EP, the Newton EP. It states ``In the Newtonian limit, the inertial and gravitational masses of a body are equal''~\cite{Casola:2015}. Since it only deals with the Newtonian limit, it is trivially obeyed in WTG. We do not see the need to analyse this EP in more detail and refer the interested readers to~\cite{Casola:2015}.

\subsection{Weak equivalence principle}
	
The weak EP reads ``Test particles with negligible self-gravity behave, in a gravitational field, independently of their properties'' \cite{Casola:2015}. By a test particle we mean one whose back-reaction on its environment can be disregarded. The negligible self-gravity requirement demands that the particle's size is much larger than the Schwarzschild radius corresponding to its mass.

A sufficient (but not necessary~\cite{Casola:2015}) condition for the EP to hold is that the effects of gravity on the trajectory of a test particle can be fully captured by the symmetric part of the connection (locally, disregarding geodesic deviations among its constituents and similar effects), which guarantees the universality of motion in a gravitational field~\cite{Casola:2015}.
	
To analyse the validity of the weak EP for WTDiff-invariant gravity, we then need to discuss motion in a gravitational field in such theories. The standard, Diff-invariant timelike geodesic equation reads
\begin{equation}
\label{geo}
u^{\nu}\nabla_{\nu}u^{\mu}=fu^{\mu},
\end{equation}
where $u^{\mu}$ denotes a unit vector tangent to the geodesic and $f=u^{\nu}\nabla_{\nu}\ln\sqrt{\vert u^2\vert}$ (for an affine parametrisation, we have $f=0$). However, this equation is not invariant under Weyl transformations. In other words, force-free trajectories in one Weyl frame are subjected to a force in a different frame. This behaviour clearly breaks the Weyl invariance of physics necessary in WTDiff-invariant gravity. 

To find a geodesic equation tailored to WTDiff-invariant gravity, we turn to one of the standard approaches to derive it in the Diff-invariant case. In particular, for any Diff-invariant theory of gravity, one can straightforwardly derive the geodesic equation for a test particle modelled by a spatially localised perfect fluid energy-momentum tensor. If the fluid is pressureless, the divergenceless condition on the energy-momentum tensor is equivalent to the geodesic equation~\eqref{geo}. As expected, the gradient of the fluid's pressure acts as a force and the particle's trajectory is no longer a geodesic.

In the WTDiff-invariant case, we consider the following WTDiff-invariant perfect fluid energy-momentum tensor $\tilde{T}_{\mu\nu}=\left(\rho+p\right)\tilde{u}_{\mu}\tilde{u}_{\nu}+p\tilde{g}_{\mu\nu}$, where the unit timelike vector $\tilde{u}^{\mu}$ is now normalised to unity with respect to the auxiliary metric. Thence, we have the following relation between $\tilde{u}^{\mu}$ and $u^{\mu}$ considered in the Diff-invariant geodesic equation
\begin{equation}
\label{WTDiff geodesic}
\tilde{u}^{\mu}=\left(\sqrt{-\mathfrak{g}}/\omega\right)^{1/D}u^{\mu}.
\end{equation}
The WTDiff-invariant divergence of this energy-momentum tensor obeys equation~\eqref{div T}
\begin{align}
\nonumber \tilde{g}^{\lambda\nu}\tilde{\nabla}_{\nu}\tilde{T}_{\lambda\mu}=& ~\tilde{g}_{\mu\lambda}\tilde{u}^{\lambda}\tilde{\nabla}_{\nu}\left[\left(\rho+p\right)\tilde{u}^{\nu}\right]\\
&+\tilde{\nabla}_{\mu}p +\left(\rho+p\right)\tilde{u}^{\nu}\tilde{\nabla}_{\nu}\tilde{u}_{\mu}=\tilde{\nabla}_{\mu}\mathcal{J},
\end{align}
where $\mathcal{J}$ is a measure of the local energy-momentum non-conservation, as we have seen above. Projecting this equation on the surface orthogonal to $\tilde{u}^{\mu}$ via the projection tensor $\tilde{h}^{\mu\rho}=\tilde{g}^{\mu\rho}+\tilde{u}^{\mu}\tilde{u}^{\rho}$ yields
\begin{equation}
\left(\rho+p\right)\tilde{u}^{\nu}\tilde{\nabla}_{\nu}\tilde{u}^{\rho}=\tilde{h}^{\rho\nu}\tilde{\nabla}_{\nu}\left(\mathcal{J}-p\right).
\end{equation}
The left hand side is proportional to the WTDiff-invariant acceleration of the test particle $\tilde{a}^{\rho}=\tilde{u}^{\nu}\tilde{\nabla}_{\nu}\tilde{u}^{\rho}$, whereas the right hand side is the force acting on the particle. Aside from the force sourced by the gradient of the pressure, there is also a new contribution sourced by the gradient of the energy non-conservation measure~$\mathcal{J}$. Therefore, a force-free trajectory of a test particle composed of a perfect fluid in WTDiff-invariant geometry is characterised by the condition
\begin{equation}
\label{force free}
\tilde{h}^{\rho\nu}\tilde{\nabla}_{\nu}\left(\mathcal{J}-p\right)=0.
\end{equation}
The equivalent condition in the Diff-invariant case reads $h^{\rho\nu}\nabla_{\nu}p=0$ since the Diff invariance implies $\mathcal{J}=0$. Since the divergenceless tensor $\tilde{T}'_{\mu\nu}$ for a perfect fluid reads
\begin{equation}
\label{fluid}
\tilde{T}'_{\mu\nu}=\left(\rho+\mathcal{J}\right)\tilde{u}_{\mu}\tilde{u}_{\nu}+\left(p-\mathcal{J}\right)\tilde{h}_{\mu\nu},
\end{equation}
its effective pressure equals precisely $p-\mathcal{J}$. Then, the condition for a force-free trajectory requires nothing but the vanishing spatial gradient of the pressure associated with the divergenceless tensor $\tilde{T}'_{\mu\nu}$. This tensor is thus the relevant measure of the energy-momentum for the purpose of identifying force-free trajectories of test particles.

For a perfect fluid obeying the condition~\eqref{force free} the divergence of the energy-momentum tensor yields the condition $\tilde{u}^{\nu}\tilde{\nabla}_{\nu}\tilde{u}^{\rho}=0$ and the particle consequently follows a timelike geodesic trajectory in any Weyl frame. Therefore, allowing for non-affine parametrisations, the appropriate WTDiff-invariant geodesic equation reads
\begin{equation}
\label{geodesic}
\tilde{u}^{\nu}\tilde{\nabla}_{\nu}\tilde{u}^{\mu}=f\tilde{u}^{\mu},
\end{equation}
where $f=\tilde{u}^{\nu}\tilde{\nabla}_{\nu}\ln\sqrt{\vert\tilde{u}^2\vert}$. It is easy to see that equation~\eqref{geodesic} yields the required WTDiff-invariant force-free trajectories. With this definition of a geodesic, any local, WTDiff-invariant theory of gravity incorporates the weak EP.

The geodesic equation~\eqref{geodesic} has further consequences for WTDiff-invariant gravity. It directly shows that, while the dynamical metric $g_{\mu\nu}$ remains the dynamical variable describing gravity, the metric relevant for describing the spacetime geometry in which matter moves is actually the auxiliary one, $\tilde{g}_{\mu\nu}$. In essence, $\tilde{g}_{\mu\nu}$ represents the kinematical metric of WTDiff-invariant gravity, as is apparent already from the fact that matter fields couple to $\tilde{g}_{\mu\nu}$. Both metrics differ only in their measure of spacetime volume, which cannot be experimentally accessed by any known method~\cite{Barcelo:2018,Carballo:2022}. Thence, using $g_{\mu\nu}$ as the dynamical variable and $\tilde{g}_{\mu\nu}$ as the way to measure distances in the spacetime does not allow us to distinguish WTDiff-invariant gravitational theories from the Diff-invariant ones on the level of classical physics~\cite{Barcelo:2018,Carballo:2022}.

A somewhat more sophisticated argument for the weak EP relies on the Geroch-Jang theorem~\cite{Geroch:1975}, which gives a useful way to characterise timelike geodesics. Let us assume that for every neighbourhood $\mathcal{U}$ of a curve $\Gamma$ there exists a tensor $\Theta_{\mu\nu}$ satisfying the following properties: 

\begin{itemize}
    \item [(i)] $\Theta_{\mu\nu}$ vanishes everywhere outside $\mathcal{U}$; 
    \item[(ii)] $\Theta_{\mu\nu}$ is nonzero somewhere in $\mathcal{U}$; 
    \item[(iii)] $\Theta_{\mu\nu}$ has vanishing divergence; and 
    \item[(iv)] $\Theta_{\mu\nu}$ satisfies the dominant energy condition, i.e., $\Theta_{\mu\nu}n^{\mu}n^{\nu}\ge0$ for every timelike vector field $n^{\mu}$ and $\Theta_{\mu\nu}n^{\nu}$ is timelike (or vanishing).
\end{itemize}
Then it follows that $\Gamma$ is a timelike geodesic.

In Diff-invariant gravity, taking $\Theta_{\mu\nu}$ to be the energy-momentum tensor $T_{\mu\nu}$ of the test particle, this theorem guarantees that the particle follows a timelike geodesic, in accord with the weak EP, provided that the energy-momentum tensor satisfies the necessary dominant energy condition.

Since applying the Geroch-Jang theorem requires that $T_{\mu\nu}$ vanishes outside of \textit{any} neighbourhood $\mathcal{U}$ of $\Gamma$, the test particle must be arbitrarily small. Of course, a more practical choice (followed also in the original proof of the theorem) is to make the body confined in a small enough radius $l$ and then systematically neglect any $O\left(l\right)$ effects. In this way, the theorem is not contradicted, e.g. by particles with nontrivial angular momentum whose motion deviate from the geodesic one at $O\left(l\right)$~\cite{Geroch:1975} (as an aside, if quantum particles with a spin were indeed fundamentally point-like, they would contribute at the order $O\left(l^0\right)$, violating the weak EP~\cite{Casola:2015}).

For WTDiff-invariant theories, one needs to apply the Geroch-Jang theorem to $\tilde{T}'_{\mu\nu}$~\eqref{T divergenceless} whose WTDiff-invariant divergence vanishes as required. Of course, demanding the dominant energy condition for $\tilde{T}'_{\mu\nu}$ rather than for $\tilde{T}_{\mu\nu}$ is a stronger requirement. However, equations~\eqref{Einstein-style} which are the divergenceless equations for WTG have $\tilde{T}'_{\mu\nu}$ on the right hand side. In other words, it plays the same role as the energy-momentum tensor $T_{\mu\nu}$ in GR. Thus, $\tilde{T}'_{\mu\nu}$ should be relevant for any application  of the energy conditions to WTDiff-invariant gravity, e.g. for the proofs of singularity theorems or for the exclusion of solutions containing closed timelike curves. As an aside, this difference is irrelevant for the null energy conditions, since $\mathcal{J}\tilde{g}_{\mu\nu}\tilde k^{\mu}\tilde k^{\nu}=0$ for any null vector  $\tilde{k}^{\mu}$. With the dominant energy condition satisfied, the Geroch-Jang theorem then ensures the validity of the weak EP for any local, WTDiff-invariant gravitational theory.

We can also heuristically understand why the weak EP should be respected by WTDiff-invariant gravity from a more general viewpoint. The weak EP endows the spacetime with a projective structure, which consists of the timelike geodesics. The projective structure together with the conformal structure, consisting of the light cones, has been shown to specify the Weyl connection (but not the Levi-Civita connection) in the spacetime~\cite{Ehlers:1972}. Both WTDiff- and Diff-invariant gravity has the same conformal structure ($g_{\mu\nu}$ and $\tilde{g}_{\mu\nu}$ are related by a conformal transformation). Moreover, the WTDiff- and Diff-invariant geodesic equations both employ a Weyl connection with respect to the dynamical metric $g_{\mu\nu}$ (for Diff-invariant theories, the connection is actually Levi-Civita). Determining this Weyl connection requires both conformal and projective structures. Since the projective structure is supplied by the weak EP, it appears to be natural that WTDiff-invariant gravity (which has a Weyl connection) obeys this principle. We leave a more mathematically precise formulation of this statement for a future work.

\subsection{Einstein equivalence principle}

A stronger condition than the weak EP is the Einstein EP\footnote{The so called Schiff's conjecture proposes that the weak and the Einstein EP are actually equivalent~\cite{Thorne:1973}, but it remains unproven~\cite{Casola:2015}. Thus, for the purposes of this work, we regard the Einstein EP as a generalisation of the weak EP.}, which extends it from the motion of particles to all non-gravitational test physics. It states that ``Fundamental non-gravitational test physics is not affected, locally and at any point of spacetime, by the presence of a gravitational field''~\cite{Casola:2015}. The status of the Einstein EP in Diff-invariant theories already presents a fairly complicated issue. In particular, the principle is limited to ``fundamental physics'' (so as to exclude, e.g. composite bodies whose behaviour can, even locally, depend on the spacetime curvature~\cite{Casola:2015}). No unambiguous way to specify the fundamental physics for the purposes of the Einstein EP has been put forward as of yet. Nevertheless, there exist criteria which a matter field theory must satisfy in order to comply with the Einstein EP~\cite{Fletcher:2022b}.

Switching from Diff-invariant to WTDiff-invariant gravity does not influence the non-gravitational test physics. In particular, Weyl transformations do not act on the matter fields and the Diff-invariant and WTDiff-invariant laws applying to non-gravitational physics thus have the same form, simply replacing $g_{\mu\nu}$ with $\tilde{g}_{\mu\nu}$ and $\nabla_{\mu}$ with $\tilde{\nabla}_{\mu}$ (see the discussion of the geodesic equation in the previous subsection). 

To make this point clearer, we can study the WTDiff-invariant formulation of one of the well-established criteria for matter fields compatible with the Einstein EP, the Ehlers criterion~\cite{Ehlers:1973}. Simply put, the Ehlers criterion requires that flat spacetime and curved spacetime solutions of the matter equations of motion are sufficiently close to each other in the vicinity of each spacetime point. A detailed evaluation of its appropriateness for choosing matter theories compatible with the Einstein EP (i.e., locally specially relativistic theories) is provided in~\cite{Fletcher:2022b} and we include a discussion of the WTDiff-invariant case in appendix~\ref{Ehlers}. For our purposes, it is crucial that the entire statement of the WTDiff-invariant Ehlers criterion (as well as of other similar criteria) indeed carries over from the Diff-invariant setup, just with the replacement $g_{\mu\nu}\to\tilde{g}_{\mu\nu}$. Hence, we clearly show that any Diff-invariant matter field theory compatible with the Einstein EP (regardless of the precise requirements for this compatibility) has an equivalent WTDiff-invariant formulation which also obeys this principle.

\subsection{Equivalence principles and local energy-momentum non-conservation}

Passing the tests for the EPs for non-gravitational test physics serves primarily as a sanity check for WTDiff-invariant gravity. Beyond that, it also illuminates the issue of the possible local energy-momentum non-conservation proposals in these theories. While we touched on this question in the previous subsections, we can now provide a synthesis of the insights from the different EPs.

To begin, we stress again that the equations of motion of WTDiff-invariant gravities depend only on the traceless part of the energy-momentum tensor. Even applying the Bianchi identities to obtain divergenceless equations, the right hand side is given by the divergenceless tensor $\tilde{T}'_{\mu\nu}$ defined by equation~\eqref{T divergenceless}. Then, the possible local energy-momentum non-conservation cannot be actually noticed at the level of the gravitational equations of motion. Then, the kinematics of matter fields in a WTDiff-invariant spacetime, constrained by the EPs, is a natural place to look for signatures of the local energy-momentum non-conservation.

In the context of the weak EP, we have shown that the relevant tensor for the Geroch-Jang theorem is again the divergenceless $\tilde{T}'_{\mu\nu}$. Likewise, the simpler derivation from the perfect fluid tensor imposes the condition of the vanishing pressure gradient on the effective pressure term $p-\mathcal{J}$, associated with the divergenceless tensor $\tilde{T}'_{\mu\nu}$, which for a perfect fluid follows equation~\eqref{fluid}.

In conclusion, the weak EP only cares about $\tilde{T}'_{\mu\nu}$. This outcome is consistent with the fact that the weak EP can be heuristically derived from the gravitational equations of motion, by considering a point-like source for them. Since the equations of motion only depend on $\tilde{T}'_{\mu\nu}$, it follows that the same applies to the weak EP.

The Einstein EP does not only concern the geodesic equation but also the equations of motion of the matter fields. Since the local energy-momentum \mbox{(non-)}conservation is determined by the matter equations of motion (assuming one can independently identify the energy-momentum tensor\footnote{This issue is in general very subtle, since theories with energy-momentum non-conservation generically lack a Lagrangian description~\cite{Josset:2017,Perez:2018,Carballo:2022}.}), the Einstein EP is sensitive to it. Nevertheless, local energy-momentum non-conservation does not signal any breaking of the Einstein EP. One only requires that the non-conservation (on the level of fundamental physics as required by the Einstein EP) does not locally depend on the spacetime curvature or on the background volume measure $\omega$. This condition only amounts to minimal coupling of the matter fields to gravity, which is a general requirement for the Einstein EP to hold. Assuming minimal coupling, the Ehlers criterion goes through by matching the solutions of the flat spacetime matter equations of motion with the curved spacetime solutions in a suitable frame. In this sense, the local energy-momentum (non-)conservation introduces no to the Einstein EP.

For instance, a perfect fluid model with energy-momentum non-conservation we applied to discuss the weak equivalence principle passes the Ehlers criterion, provided that $\mathcal{J}$ does not depend on the Riemann tensor or on $\omega$. It is easy to see that the flat spacetime and curved spacetime solutions can be matched in this case. A special case of such a perfect fluid is the continuous spontaneous localisation model~\cite{Ghirardi:1986,Pearle:1989} suggested in the cosmological context as a possible source for the acceleration of the universe~\cite{Josset:2017}. This proposal involves no violation of the Einstein EP on the level of classical physics\footnote{A later proposal of energy dispersion due to granular structure of the spacetime~\cite{Perez:2018} represents a special case, as it involves point-like spinning particles. These are known to generically violate even the weak EP~\cite{Casola:2015}. The EP violation then does not depend on the local energy non-conservation occurring in this model.}.

\section{Gravitational equivalence principles}
\label{G EP}

The weak EP fixes the kinematics of motion in a gravitational field, essentially requiring that the trajectories of the test particles are determined by the connection. The Einstein EP further specifies the behaviour of matter fields in a curved background. However, they are both compatible with essentially any local, Diff- or WTDiff-invariant theory of gravity and do not really constrain the gravitational dynamics in any way. For that, some version of the EP which applies to test gravitational physics is necessary. These EPs present far more stringent constraints and only the Diff-invariant Lanczos-Lovelock theories of gravity are known to be compatible even with the gravitational weak EP. In this section, we show that WTG (as well as the WTDiff-invariant Lanczos-Lovelock theories) also obeys the gravitational weak EP and that WTG incorporates even the more stringent strong EP.

\subsection{Gravitational weak equivalence principle}
\label{GWEP main}
	
The weak EP can also be  generalised to apply to self-gravitating test particles. The resulting formulation is known as the  gravitational weak EP which asserts that  ``Test particles behave, in a gravitational field and in vacuum, independently of their properties''~\cite{Casola:2015}. The gravitational weak EP does not represent a direct generalisation of the weak EP, since it is restricted to vacuum. More precisely, we require that there exists a sufficiently large region around the test particle (much larger than its Schwarzschild radius) in which no matter is present.  Otherwise, the intrinsic gravitational field of the test particle would influence the nearby matter, thus breaking the universality.
	
A simple criterion for the validity of the gravitational weak EP utilises the Geroch-Jang theorem~\cite{Casola:2014}. However, rather than applying the theorem just to the energy-momentum tensor of the test particle, it also needs to include the perturbation of the gravitational field caused by the presence of the particle (i.e., the effective energy-momentum of its gravitational field). Moreover, one must keep in mind that the geodesic along which the test particle should move lies in the unperturbed spacetime. Splitting the WTDiff-invariant auxiliary metric into the background part $\tilde{g}_{\mu\nu}$ and the perturbation caused by the test particle, $\tilde{\gamma}_{\mu\nu}$, we may similarly split the equations of motion. In the case of WTG, we obtain the vacuum divergenceless equations for the background metric
\begin{equation}
\tilde{G}_{\mu\nu}=\Lambda\tilde{g}_{\mu\nu},
\end{equation}
and the equations governing the perturbation\footnote{A subtle issue should be noted. In WTDiff-invariant gravity, the perturbation in principle also changes the value of the cosmological constant, which is an on-shell integration constant. However, it does not seem realistic that a test particle of infinitesimal size should change the global value of the cosmological constant, as the equations of motion would then require a corresponding global change in the spacetime curvature. Then, the gravitational effect of the test particle would no longer be localised, breaking one of the assumptions under which the gravitational weak EP can be expected to hold. Therefore, we set $\delta\Lambda=0$ in the following.}
\begin{equation}
\label{perturbation}
\tilde{\mathcal{G}}_{\mu\nu}+\Lambda\tilde{\gamma}_{\mu\nu}=8\pi G\left(\tilde{T}'_{\mu\nu}-\tilde{T}^{\text{(g)}}_{\mu\nu}\right)\equiv 8\pi G\tilde{\mathcal{T}}_{\mu\nu},
\end{equation}
where $\tilde{\mathcal{G}}_{\mu\nu}$ denotes the perturbation of the WTDiff-invariant auxiliary Einstein tensor. The first term on the right hand side $\tilde{T}'_{\mu\nu}$ corresponds to the divergenceless energy-momentum tensor of the test particle. The second term $\tilde{T}^{\text{(g)}}_{\mu\nu}$ quantifies the effective (WTDiff-invariant) energy-momentum of the gravitational field, which is quadratic in the auxiliary metric perturbation $\tilde{\gamma}_{\mu\nu}$\footnote{Naturally, one actually perturbs the dynamical metric $g_{\mu\nu}$, $\tilde{\gamma}_{\mu\rho}$ is simply a convenient book-keeping device.}. The tensor $\tilde{\mathcal{T}}_{\mu\nu}$ then quantifies both the energy-momentum of the test particle and its gravitational self-energy.

The tensor $\tilde{\mathcal{T}}_{\mu\nu}$ satisfies the conditions of the Geroch-Jang theorem with respect to the background (unperturbed) metric. Indeed, conditions (i) and (ii) concerning the localisation of the tensor are trivial. Validity of the dominant energy condition (condition (iv)) represents a nontrivial assumption, but it is satisfied for ``reasonable'' test particles~\cite{Casola:2014}. In fact, if condition (iv) did not hold, $\tilde{\mathcal{G}}_{\mu\nu}$ could violate the timelike geodesic focusing theorem (and the singularity theorems which follow from it), as it enters the Raychaudhuri equation~\cite{Raychaudhuri:1955,Hawking:1973}.

Lastly, we must check condition (iii), i.e., that $\tilde{\nabla}_{\nu}\tilde{\mathcal{T}}_{\mu}^{\;\:\nu}=0$, where the covariant derivative $\tilde{\nabla}_{\nu}$ is defined with respect to the background metric $\tilde{g}_{\mu\nu}$. The gravitational energy-momentum $\tilde{T}^{\text{(g)}}_{\mu\nu}$ is a complicated expression quadratic in the metric perturbation $\tilde{\gamma}_{\mu\nu}$. It is then more convenient to check that $\tilde{\nabla}_{\nu}\tilde{\mathcal{G}}_{\mu}^{\;\:\nu}=0$ and use $\tilde{\mathcal{G}}_{\mu\nu}=8\pi G\tilde{\mathcal{T}}_{\mu\nu}$ thanks to equation~\eqref{perturbation}~\cite{Casola:2014}.

The divergenceless condition then reads \mbox{$\tilde{\nabla}^{\nu}\left(\tilde{\mathcal{G}}_{\mu\nu}+\Lambda\tilde{\gamma}_{\mu\nu}\right)=0$}. Writing it in terms of the metric perturbation $\tilde{\gamma}_{\mu\nu}$ and its derivatives, we obtain
\begin{align}
\nonumber
&\tilde{\nabla}^{\nu}\left(\tilde{\mathcal{G}}_{\mu\nu}+\Lambda\tilde{\gamma}_{\mu\nu}\right)=\frac{1}{2}\tilde{\nabla}^{\nu}\Big[2\tilde{\nabla}^{\lambda}\tilde{\nabla}_{(\mu}\tilde{\gamma}_{\nu)\lambda} \\
\nonumber &-\tilde{\nabla}^{\lambda}\tilde{\nabla}_{\lambda}\tilde{\gamma}_{\mu\nu}+\tilde{g}_{\mu\nu}\left(-\tilde{\nabla}_{\lambda}\tilde{\nabla}_{\rho}\tilde{\gamma}^{\lambda\rho}+\tilde{R}_{\lambda\rho}\tilde{\gamma}^{\lambda\rho}\right) \\
&-\tilde{R}\tilde{\gamma}_{\mu\nu}+2\Lambda\tilde{\gamma}_{\mu\nu}\Big] \label{perturbation 2},
\end{align}
where $\tilde{g}_{\mu\nu}$, $\tilde{\nabla}_{\mu}$, $\tilde{R}_{\mu\nu}$ denote the corresponding background quantities. We can simplify equation~\eqref{perturbation 2} by commuting the derivatives and using the definition of the auxiliary Riemann tensor. Then, we obtain
\begin{align}
\nonumber &\tilde{\nabla}^{\nu}\left(\tilde{\mathcal{G}}_{\mu\nu}+\Lambda\tilde{\gamma}_{\mu\nu}\right)=\tilde{R}_{\mu}^{\;\:\lambda}\tilde{\nabla}_{\nu}\tilde{\gamma}^{\nu}_{\lambda}+\tilde{\gamma}^{\nu}_{\lambda}\tilde{\nabla}_{\nu}\tilde{R}_{\mu}^{\;\:\lambda} \\
&+\frac{1}{2}\tilde{R}^{\;\:\lambda}_{\nu}\tilde{\nabla}_{\mu}\tilde{\gamma}^{\nu}_{\lambda}-\frac{1}{2}\tilde{R}_{\mu\nu}\tilde{\nabla}^{\nu}\tilde{\gamma}-\frac{1}{2}\tilde{\nabla}_{\nu}\left(\tilde{R}\tilde{\gamma}^{\nu}_{\mu}\right)+\Lambda\tilde{\nabla}_{\nu}\tilde{\gamma}^{\nu}_{\mu}. \label{GWEP test 2}
\end{align}
Equations~\eqref{Einstein-style} applied to the vacuum background allow us to write the Ricci tensor in terms of $\Lambda$, i.e. $\tilde{R}_{\mu\nu}=2\Lambda\tilde{g}_{\mu\nu}/\left(D-2\right)$. In addition,  $\tilde{\gamma}=\tilde{g}^{\mu\nu}\tilde{\gamma}_{\mu\nu}=0$, since the determinant of the auxiliary metric is fixed to $\omega$. Indeed, in terms of the perturbation $\gamma_{\mu\nu}$ of the dynamical metric, it holds $\tilde{\gamma}_{\mu\nu}=\left(\sqrt{-\mathfrak{g}}/\omega\right)^{-2/D}\left(\gamma_{\mu\nu}-g_{\mu\nu}\gamma/D\right)$. Thence, $\tilde{\gamma}_{\mu\nu}$ is traceless. This allows us to conclude that the right hand side of equation~\eqref{GWEP test 2} indeed vanishes identically.

In total, $\tilde{\mathcal{T}}_{\mu\nu}$ satisfies all the conditions of the Geroch-Jang theorem. It follows that the test particle moves along a timelike geodesic of the unperturbed auxiliary metric and, consequently, the gravitational weak EP holds. We have shown that, just like GR, WTG incorporates the gravitational weak EP. 

Regarding the more general WTDiff-invariant theories, only Lanczos-Lovelock gravity obeys the gravitational weak EP. The proof would be a simple modification of the argument presented for Diff-invariant gravity~\cite{Casola:2014}, which has reached the same conclusion. In summary, WTG and GR are the only two metric gravitational theories in four dimensions known to be compatible with the gravitational weak EP (Nordström gravity also satisfies this principle, but it is not metric and, therefore, incompatible with the Einstein EP~\cite{Casola:2014}).

\subsection{Strong equivalence principle}

Lastly, the strong EP extends the Einstein EP to include test gravitational physics: ``All test fundamental physics (including gravitational physics) is not affected locally by the presence of a gravitational field''~\cite{Casola:2015}. It relates to the Einstein EP in an analogous way as the gravitational weak EP does to the weak EP, extending it to include gravitational fields with a negligible back reaction on the background spacetime. The strong EP can also be phrased as the requirement of local Poincar\'{e} invariance of all the test physics, including gravitational physics (e.g. the local behaviour of linearised gravitational waves on a curved background), combined with the validity of the gravitational weak EP~\cite{Casola:2015}. In the previous subsection, we have proven the latter requirement for WTG. We have also argued in section~\ref{NG EP} that the condition of local Poincar\'{e} invariance of matter test physics can be implemented in the WTDiff-invariant setting without any issues (via, e.g. Ehlers criterion). The only remaining issue is the local Poincar\'{e} invariance of gravitational test physics, in particular, in regards to weak gravitational waves. However, linearised WTG is manifestly Poincar\'{e} invariant~\cite{Barcelo:2014,Barcelo:2018,Carballo:2022}, implying local Poincar\'{e} invariance of gravitational test physics. The strong EP then does apply to WTG, as it satisfies all the necessary conditions. Consequently, GR and WTG seem to be the only two known gravitational theories in $4$ spacetime dimensions compatible with the strong EP. The uniqueness of these two theories can be traced to their status as the only two consistent theories of massless, self-interacting gravitons with no additional degrees of freedom~\cite{Barcelo:2018} (or gauge fixing). Any other propagating gravitational degrees of freedom tend to destroy the gravitational weak EP~\cite{Casola:2014}. For instance, if a non-minimally coupled scalar field is present, the geodesic motion of self-gravitating test particles depends on its local value, breaking the gravitational weak EP.

\section{Conclusions}
\label{conclusion}

We have explored the validity of various formulations of the equivalence principle in WTDiff-invariant theories and, in particular, in WTG. Our discussion shows how the EPs serve to illuminate the main features of gravity and its coupling to matter. That becomes especially useful in the case of WTDiff-invariant gravity, whose local symmetries and dependence on a nondynamical volume measure lead to behaviour which differs from the usual intuition one develops for Diff-invariant theories. 

The weak EP exposes why the geodesic equation needs to be written in term of the Weyl connection $\tilde{\Gamma}^{\mu}_{\;\:\nu\rho}$. Moreover, it points towards the appropriate way for handling the potential local energy-momentum non-conservation in WTG. The Geroch-Jang theorem which implies the weak EP takes as its input the divergenceless tensor $\tilde{T}'_{\mu\nu}$. This tensor also appears as the right hand side of the divergenceless equations for gravitational dynamics. Therefore, it completely determines the motion of matter in a gravitational field. The equations of motion for the matter fields can  be further derived from the condition $\tilde{\nabla}^{\nu}\tilde{T}'_{\mu\nu}=0$. Then, on the level of the gravitational dynamics, $\tilde{T}'_{\mu\nu}$ fulfills all the roles of a locally conserved energy-momentum tensor. Nevertheless, the non-conserved energy-momentum tensor $\tilde{T}_{\mu\nu}$ can still be significant if it corresponds to the definitions of energy, momenta and stresses that are ``natural'' from the perspective of the matter field dynamics (see, e.g.~\cite{Ghirardi:1986,Pearle:1989} for an example of such a matter theory).

Similarly, the Einstein EP shows that any Diff-invariant theory of matter fields can be rewritten to a WTDiff-invariant one by simply replacing $g_{\mu\nu}$ with $\tilde{g}_{\mu\nu}$ and $\nabla_{\mu}$ with $\tilde{\nabla}_{\mu}$. The resulting theory continues to satisfy the Einstein EP, i.e., it remains locally special relativistic.

The EPs for gravitational test physics finally restrict the dynamics. The gravitational weak EP already restricts us to Lanczos-Lovelock theories of gravity (both Diff- and WTDiff- invariant), together with non-metric Nordström gravity~\cite{Casola:2014}. The former has been shown to fail the strong EP~\cite{Alonso:2025}. The latter is also trivially ruled out by the strong EP. It essentially amounts to requiring that the Einstein EP holds and that the only propagating degrees of freedom of the theory are those associated with a massless graviton. In $4$ spacetime dimensions, the only two theories compatible with the strong EP are then GR and WTG.

The potential of employing the EPs to classify and better understand various candidate theories of gravity, as outlined in Refs.~\cite{Thorne:1973,Casola:2014,Casola:2015}, is of course not limited to the WTDiff-invariant case. Even in the (broadly defined) area of unimodular gravity, the same concept can be applied, e.g. to the recently proposed Diff-invariant first order formulation~\cite{Montesinos:2023,Montesinos:2025}, to the gauge-fixed theory~\cite{Padilla:2014}, to the teleparallel version~\cite{Nakayama:2023}, and to WTDiff-invariant gravity with dynamical volume $D$-form~\cite{Carballo:2020}. The EPs could help establish a hierarchy of these approaches, whose physical (in)equivalence remains somewhat obscure in the literature. Our work may then also serve as a first step in this direction.

\section*{Acknowledgments}

AA-S is funded by the Deutsche Forschungsgemeinschaft (DFG, German Research Foundation) — Project ID 51673086. ML is supported by the DIAS Post-Doctoral Scholarship in Theoretical Physics 2024 and by the Charles University Grant Agency project No. GAUK 90123. AA-S and LJG acknowledge support through Grant  No.  PID2023-149018NB-C44 (funded by MCIN/AEI/10.13039/501100011033). LJG also acknowledges the support of the Natural Sciences and Engineering Research Council of Canada (NSERC).

\appendix

\section{The WTDiff-invariant Ehlers criterion}
\label{Ehlers}

In this appendix, we provide a WTDiff-invariant formulation of one of the well-established criteria for matter fields compatible with the Einstein EP, the Ehlers criterion~\cite{Ehlers:1973} (using its ``g-approximating'' version proposed in~\cite{Fletcher:2022b}). For a WTDiff-invariant theory, the criterion can be stated in the following way. We start with a manifold $M$ endowed with a dynamical metric $g_{\mu\nu}$ and a nondynamical volume $D$-form $\boldsymbol{\omega}$ (which together allow us to construct the auxiliary metric $\tilde{g}_{\mu\nu}$). We further consider a collection of fields $\psi$. We prescribe the equations of motion for the matter fields to be $A[\tilde{g}_{\mu\nu},\psi]=0$, where $A[\tilde{g}_{\mu\nu},\psi]$ is a differential expression constructed from the auxiliary metric $\tilde{g}_{\mu\nu}$, the fields $\psi$, and their derivatives. In a neighbourhood $O$ of an arbitrary regular point $P\in M$ we introduce a flat spacetime metric $\eta_{\mu\nu}$ such  that it agrees with $\tilde{g}_{\mu\nu}$ in any point $Q\in O$ to the first order in the geodesic distance between $P$ and $Q$. Consider any $\bar{\psi}$ that solves the equation $A[\eta_{\mu\nu},\bar{\psi}]=0$ in $O$. Then, for any $\epsilon>0$, there exists some $\psi$ solving the curved spacetime equation $A[\tilde{g}_{\mu\nu},\psi]=0$ in some neighborhood $O'$ of $P$, such that $O'\subseteq O$, which approximates $\bar{\psi}$. More precisely, we require that there exists some neighborhood $U$ of $P$, such that $U\subseteq O'$ and $\vert\vert\bar{\psi}-\psi\vert\vert<\epsilon$ in $U$. The choice of a suitable norm $\vert\vert\cdot\vert\vert$ depends on the context, one possible example being the $C^k$ supremum norms on the sub-neighborhood~$U$~\cite{Fletcher:2022b}.

If a matter field theory satisfies the Ehlers criterion, it is locally specially relativistic in the following sense. A solution to the flat spacetime matter equations of motion is approximated by some curved spacetime solution in the neighborhood of any given point and this approximation is arbitrarily good in a sufficiently small neighborhood. It is easy to see that this statement indeed implies the validity of the Einstein EP for the given matter field coupled to a metric, Diff- or WTDiff-invariant theory of gravity.

\bibliography{bibliography}

\end{document}